\documentclass[prd,twocolumn, 10pt]{revtex4-1}

\usepackage{bm}
\usepackage{epsfig}
\usepackage{amsmath}
\usepackage{latexsym}
\usepackage{slashed} % for Dirac slash
\usepackage{mathrsfs}
\usepackage{graphicx}
\usepackage{color}

\begin{document}

\title{{\large QCD Flux Tubes and Anomaly Inflow}}

\author{Chi Xiong}
\email[]{xiongchi@ntu.edu.sg}
\affiliation{~~~~~~\\Institute of Advanced Studies, Nanyang Technological University, \\
     Singapore 639673 }

%\date{\today}

\begin{abstract}

We apply the Callan-Harvey anomaly inflow mechanism to the study of QCD  (chromoelectric) flux tubes, quark (pair)-creation and chiral magnetic effect, using new variables from the Cho-Faddeev-Niemi decomposition of the gauge potential. A phenomenological description of chromoelectric flux tubes is obtained by studying a gauged Nambu-Jona-Lasinio effective Lagrangian, derived from the original QCD Lagrangian. At the quantum level, quark condensates in the QCD vacuum may form a vortex-like structure in a chromoelectric flux tube. Quark zero modes trapped in the vortex are chiral and lead to a two-dimensional gauge anomaly. To cancel it an effective Chern-Simons coupling is needed and hence a topological charge density term naturally appears. 

\vspace{0.5cm}


\end{abstract}

\maketitle               
                
\section{Introduction}

Flux tubes in quantum chromodynamics (QCD) play important roles in many interesting places like confinement, quark pair creation, hadron structure and phase transitions. In the dual-superconductor description of QCD vacuum \cite{'tHooft}, the chromoelectric field lines between color sources, like a quark and antiquark pair, are squeezed into a narrow flux tube along the line connecting the pair. The resulting potential is linear and consequently leads to color confinement.  On the other hand, these flux tubes can provide chromoelectric field strong enough for quark pair creation, via the usual Schwinger mechanism \cite{Schwinger} (note that this is a non-perturbative particle creation mechanism) \cite{Casher}. QCD flux tubes are also well-studied in the formation of quark gluon plasma (QGP) in heavy-ion collisions. 

Another subject of long interest, is the topological charge and its distribution in the QCD vacuum. Different topological defects in the QCD vacuum have been explored by large-$N_c$ methods, holographic QCD and lattice simulations. Among the topological defects vortex is particularly interesting, especially when we consider its connection to the flux tubes and how it interacts with the quarks. For example, the localization of quark zero modes on the topological defects leads to an interesting phenomenon --- anomaly inflow \cite{Callan-Harvey} which could shed some light on the study of topological charge distribution in the QCD vacuum. It has been shown \cite{Callan-Harvey, Naculich} that chiral fermion zero modes are localized on an axion string or a domain-wall embedded in a higher dimensional spacetime. When the fermions are coupled to some external gauge potential, gauge anomaly appears on the string or domain-wall and it is cancelled by the gauge variation of an effective action, a Chern-Simons type coupling living in the bulk of the higher dimensional spacetime.

The purpose of this paper is to provide a new description for the QCD flux tubes and to study how topological charge emerges via the anomaly inflow mechanism. Our formulation is based on a new set of variables in non-Abelian gauge theories (Cho-Faddeev-Niemi) \cite{Cho, Faddeev, DuanGe}: The usual gauge potential $A_{\mu}$ splits into two new gauge potentials $\tilde{A}_{\mu}$ and $B_{\mu}$, which are quite convenient and efficient in studying both confinement and chiral symmetry breaking, since this decomposition reveals the Abelian sector of QCD and the associated topological defects in a gauge-independent way.
As it has been shown in Refs. \cite{Cho:1999ar, Kondo:2008su}, the gauge potential $\tilde{A}_{\mu}$ alone is responsible for the Wilson loop and the Polykov loop at the operator level, while the potential $B_{\mu}$ can be integrated out in the presence of dynamical quarks to produce a nonlocal four-fermion Nambu-Jona-Lasinio (NJL) interaction, which can be used to study chiral symmetry breaking as an effective Lagrangian method \cite{Hatsuda}.  
This decomposition and the resultant physics become more complicated in our formulation, since the chromoelectric fields in the flux tube can cause a phase transition -- In fact if the flux-tube field lines are strong enough the chiral symmetry can be restored \cite{Suganuma}. Therefore the quark condensates inside the flux tube vanishes, i.e. $\langle \bar{q}q \rangle = 0 $, while outside the flux tube the QCD vacuum has non-vanishing $\langle \bar{q}q \rangle \neq 0 $. Furthermore if this configuration carries a topological charge, the quark condensates can form a vortex or even vortex lattice, similar to the Abrikosov lattice in type-II superconductors \cite{Abrikosov:1956sx}. Therefore a strong chromoelectric flux tube can induce vortices in the quark condensates.  We then can show that the chiral anomaly naturally emerges in the bulk as a Chern-Simons effective action in the anomaly-inflow scenario. When chromomagnetic fields are included, this formulation can be used to describe the chiral magnetic effect in QGP \cite{Fukushima08, Kharzeev07}. While selecting relevant degrees of freedom to study the flux tubes and enjoying the simplicity of calculations due to its Abelian character, this formulation connects different phenomena, such as confinement, chiral symmetry breaking and quark creation, to the original gauge potential and the QCD Lagrangian.

This paper is arranged as follows: In Sec. II we introduce the Cho-Faddeev-Niemi decomposition of non-Abelian gauge theories and show how the original QCD Lagrangian splits consequently; Sec. III has a brief review of the Callan-Harvey anomaly-inflow mechanism; We then study chromoelectric flux tubes in details in Sec IV, where we explain why vortices in the quark condensates can be induced by the flux tube and how anomaly and topological charges come into play; As an application chiral magnetic effect and quark creation are considered in Sec. V; Discussions and conclusions are given in Sec. VI.

\section{Cho-Faddeev-Niemi Decomposition}

In this section we briefly review the reformulation of a non-Abelian gauge theory in terms of a set of new variables, or the Cho-Faddeev-Niemi decomposition \cite{Cho, Faddeev, DuanGe, Shabanov, Kondo, Cho2}. This formulation provides a new description for the QCD flux tube as it will be shown in Sec. IV. We consider the SU(2) case for simplicity and the generalization to SU(3) or SU(N) can be found in \cite{Faddeev, Shabanov, Kondo:2008su}. This decomposition gives an Abelian projection and the associated topological structures in a gauge-invariant way, and as shown in \cite{Kondo} it is also quite convenient in studying the crossover of the confinement and chiral symmetry breaking and the topological feature of the gauge configurations. The decomposition reads  \cite{Cho, Faddeev, DuanGe, Shabanov, Kondo, Cho2}
\begin{equation} \label{CFN}
A_{\mu} = \tilde{A}_{\mu} + B_{\mu}
\end{equation}
where the new variables $\tilde{A}_{\mu}$ and $B_{\mu}$ are defined as
\begin{eqnarray}
\tilde{A}_{\mu} &=& (A_{\mu} \cdot {\bf n}) {\bf n} + i g^{-1} [ {\bf n}, \partial_{\mu} {\bf n} ] \cr
B_{\mu} &=& i g^{-1} [ \nabla_{\mu} {\bf n}, {\bf n}]
\end{eqnarray}
with $\nabla_{\mu} {\bf n} = \partial_{\mu} {\bf n} - i \, g [A_{\mu}, {\bf n}]$ and a unit vector ${\bf n}$ in the color space, satisfying  
\begin{equation}
{\bf n} \cdot {\bf n} =1.
\end{equation}
Under the gauge transformation $ \tilde{A}_{\mu}$ transforms like the original gauge field $A_{\mu}$, while $B_{\mu}$ and ${\bf n}$ transform covariantly in the adjoint representation. 
One can define a field strength from $\tilde{A}_{\mu}$
\begin{equation}
\tilde{F}_{\mu\nu} = \partial_{\mu} \tilde{A}_{\nu} - \partial_{\nu} \tilde{A}_{\mu} - i \,g [\tilde{A}_{\mu}, \tilde{A}_{\nu}]
\end{equation}
which depends on $A_{\mu}$ and ${\bf n}$ as
\begin{eqnarray}
\tilde{F}_{\mu\nu} &=& G_{\mu\nu} {\bf n}, \\ \nonumber
G_{\mu\nu} & \equiv &  \partial_{\mu} (A_{\nu} \cdot {\bf n})  - \partial_{\nu} ( A_{\mu} \cdot {\bf n}) + i g^{-1} {\bf n} \cdot [\partial_{\mu} {\bf n}, \partial_{\nu} {\bf n}]. 
\end{eqnarray}  
The original field strength can be decomposed as well
\begin{equation}
F_{\mu\nu} = F^{\parallel}_{\mu\nu} + F^{\perp}_{\mu\nu}
\end{equation}
where $F^{\parallel}_{\mu\nu}$ and $F^{\perp}_{\mu\nu}$ are defined as
\begin{eqnarray}
F^{\parallel}_{\mu\nu} &=& (F_{\mu\nu} \cdot {\bf n} ) {\bf n}, \cr 
F^{\perp}_{\mu\nu} &=& \tilde{\nabla}_{\mu} B_{\nu} - \tilde{\nabla}_{\nu} B_{\mu},  
\end{eqnarray}
with $\tilde{\nabla}_{\mu} = \partial_{\mu} - i \, g[\tilde{A}_{\mu}, ~~]$. To see the Abelian character of this decomposition, we follow Ref. \cite{Cho2} and introduce a basis $({\bf n}^1, {\bf n}^2, {\bf n}^3)$ in the $SU(2)$ color space. The color vector ${\bf n}$ can be identified with ${\bf n}^3$. Then the gauge potential $\tilde{A}_{\mu}$ can be rewritten as
\begin{equation} \label{U1}
\tilde{A}_{\mu}  = \Omega^{\textrm{\tiny{vac}}}_{\mu} + ( C_{\mu} + H^3_{\mu}) {\bf n}^3
\end{equation}
where $\Omega^{\textrm{\tiny{vac}}}_{\mu}$ represents a classical QCD vacuum 
\begin{equation}
\Omega^{\textrm{\tiny{vac}}}_{\mu} \equiv - H^a_{\mu} {\bf n}^a
\end{equation}
since the corresponding field strength vanishes. The chromomagnetic potential $H^a_{\mu}$ is defined by
\begin{equation}
H^a_{\mu} \equiv - \frac{1}{2g} \epsilon^{abc}  {\bf n}^b \partial_{\mu} {\bf n}^c
\end{equation}
and the chromoelectric potential $C_{\mu} = A_{\mu} \cdot {\bf n}^3$.  The field strength $\tilde{F}_{\mu\nu}$ becomes
\begin{equation} \label{CHmunu}
\tilde{F}_{\mu\nu} = ( C_{\mu\nu} + H_{\mu\nu}) {\bf n}^3,
\end{equation}
where the chromoelectric field strength $C_{\mu\nu}$ and the chromomagnetic field strength $H_{\mu\nu}$ are given by 
\begin{equation} \label{Cmunu}
C_{\mu\nu} =  \partial_{\mu} C_{\nu} - \partial_{\nu} C_{\mu} ,
\end{equation}
\begin{equation} \label{Hmunu}
H_{\mu\nu} =  \partial_{\mu} H^3_{\nu} - \partial_{\nu} H^3_{\mu} + g^{-1} {\bf n}^1 \cdot [\partial_{\mu},  \partial_{\nu}] {\bf n}^2
\end{equation}
respectively.  The last term in (\ref{Hmunu}) vanishes except for the cases like the Dirac string associated to a magnetic monopole. From equations (\ref{U1} -- \ref{Hmunu}) we see the Abelian nature of the potential $\tilde{A}_{\mu}$, which is crucial in studying the confinement problem as in the usual Abelian projection approach \cite{'tHooft}. This approach and other Abelianized methods have frequently been used in the flux-tube models (see for example, refs \cite{DiGiacomo, Suganuma, Tanji:2010eu}). One of the advantages in using the decomposition (\ref{CFN}) is that it separates the $U(1)$ component of the original connection in a gauge-independent way. Therefore we are not baffled by the question whether we are modelling a physical object or merely a gauge artefact.
More importantly, it has been shown in \cite{Cho:1999ar, Kondo:2008su}, the gauge potential $\tilde{A}_{\mu}$ alone is responsible for the Wilson loop and the Polykov loop at the operator level. These motivate us to use the component $\tilde{A}_{\mu}$ to describe the flux tube which is supposed to be responsible for confinement.    

The action of pure Yang-Mills theory then splits into
\begin{eqnarray}
S_{\textrm{YM}} &=& \int d^4 x \big[ - \frac{1}{4} F_{\mu\nu}^2 \big] \cr 
&=& \int d^4 x \bigg[ - \frac{1}{4} \big[ G_{\mu\nu}^2 + B_{\mu\nu}^2 + 2 (\tilde{\nabla}_{\mu} B_{\nu})^2  \cr
&& + 4 ( G_{\mu\nu} {\bf n} + \tilde{\nabla}_{\mu} B_{\nu}) \cdot B_{\mu\nu} \big] \bigg]  
\end{eqnarray}
where $B_{\mu\nu} \equiv - i \,g [B_{\mu}, B_{\nu}] $ and we have used the reduction condition \cite{Shabanov, Kondo}
\begin{equation} \label{reduction}
\tilde{\nabla}_{\mu} B^{\mu} = 0.
\end{equation}
This condition also makes the dynamical degrees of freedom of the new variables match with those of original gauge potential. It is useful to rewrite the action in the form of  \cite{Cho, Faddeev, Shabanov, Kondo}
\begin{equation}
S_{\textrm{YM}} = \int d^4 x \big[ - \frac{1}{4} G_{\mu\nu}^2  - \frac{1}{4} B_{\mu\nu}^2 -\frac{1}{2} B^{\mu} Q_{\mu\nu} B^{\nu} \big]   
\end{equation}
with the operator $Q_{\mu\nu}$ defined as
\begin{equation}
Q_{\mu\nu}^{ij} \equiv - \eta_{\mu\nu} (\tilde{\nabla}_{\tau} \tilde{\nabla}^{\tau})^{ij} + 2 g \epsilon^{ijk} \tilde{F}^k_{\mu\nu} .
\end{equation}
The quark sector of the QCD action in terms of the new variables will be given and studied in Sec.IV where we show that a {\it gauged} non-local Nambu-Jona-Lasinio effective action is obtained after integrating our the $B_{\mu}$ field, so we can study the chiral symmetry breaking as well. This reflects the convenience of using the decomposition (\ref{CFN}) in the crossover of confinement and chiral symmetry breaking in QCD.  

%%%%%%%%%%%%%%%%%%%%%%%%%%%%%%%%%%%%%%%%%%%
\section{Anomaly Inflow Mechanism}
%%%%%%%%%%%%%%%%%%%%%%%%%%%%%%%%%%%%%%%%%%%%

In the flux tube picture the chromoelectric field between color sources can be strong enough for quark creation. If these quarks and anti-quarks are created in a topologically nontrivial background, say vortices or domain-walls, their zero modes will basically live on the topological defects  \cite{Callan-Harvey}.
The localization of chiral fermion zero modes on vortices or domain-walls brings an interesting topic --- anomaly inflow mechanism \cite{Callan-Harvey} into the study of topological charge distribution in the QCD vacuum.
It has been shown \cite{Callan-Harvey} that chiral fermion zero modes are localized on an axion string or a domain-wall embedded in a higher dimensional spacetime. When the chiral zero modes coupled to some external gauge potential, gauge anomaly appears on the string or domain-wall and it is cancelled by the gauge variation of an effective action, a Chern-Simons type coupling living in the higher dimensional spacetime.
In general, chiral anomaly in $(2n+2)$-dimensions specifies the coupling between the gauge fields and the topological background. This coupling leads to the net flux of gauge charge into the topological defect, cancelling the gauge anomaly due to the chiral zero-mode fermions in $2n$-dimensions. This is how the chiral anomaly in $(2n+2)$-dimensions is related to the gauge anomaly in $2n$-dimensions. From a geometric point of view, they are related by the Stora-Zumino descent equations \cite{Zumino83, Stora83}. 

Let us take Callan and Harvey's axion-string model as an example \cite{Callan-Harvey, Naculich}--- an axion string, described by a complex scalar field $\Phi = f(\rho) e^{i \theta}$ ($\rho, \theta$ are the polar coordinates on the plane transverse to the string), is embedded in a four-dimensional bulk space. Fermions coupled to the string have chiral zero modes localized on it. If the fermions are also coupled to some gauge potential, there will be a two-dimensional gauge anomaly due to the fermion zero modes
\begin{equation}
D^k J_k = \frac{1}{2\pi} \epsilon^{ij} \partial_i A_j, ~~~~~~i,j,k=0,1
\end{equation}
which can only be cancelled by a Chern-Simons type coupling  \cite{Callan-Harvey}
\begin{equation} \label{axionstring}
S_{\textrm{\tiny{eff}}} = - \frac{1}{8 \pi^2} \int d^4x \, \partial_{\mu} \theta \, K^{\mu}.
\end{equation}   
Note that there is  a subtlety in using covariant anomaly or consistent anomaly and it has been clarified in \cite{Naculich}. Here we use the covariant anomaly. The effective coupling
\begin{equation} \label{CSphi}
\mathcal{L_{\textrm{\tiny{int}}}} = \partial_{\mu}\theta \, K^{\mu}
\end{equation}
contains the Chern-Simons current
\begin{equation}
K_{\mu} = \epsilon_{\mu\alpha\beta\gamma} \textrm{Tr} (A^{\alpha} \partial^{\beta} A^{\gamma} + \frac{2}{3} A^{\alpha} A^{\beta} A^{\gamma})
\end{equation}
which comes from the Chern-Simons form $ \mathcal{K}_{cs} = \textrm{Tr} (A \wedge F - \frac{1}{3} A \wedge A \wedge A ) $.
Integrated by parts it gives a term resembling the QCD $\theta$-term 
\begin{equation} \label{TC}
\theta \, F_{\mu\nu} \tilde{F}^{\mu\nu}
\end{equation}
with the help of the identity
\begin{equation}
d\, \mathcal{K}_{cs} = \textrm{Tr} ~F \wedge F.
\end{equation}
Note that the topological charge density in (\ref{TC}) is gauge invariant, while the Chern-Simons current coupling (\ref{CSphi}) is not. This is because the Chern-Simons form transforms under the gauge transformation
\begin{eqnarray}
A &\longrightarrow & A' = g^{-1} A g + g^{-1} dg, \cr
\mathcal{K}_{cs} & \longrightarrow & \mathcal{K}'_{cs} = \mathcal{K}_{cs} - d\,\textrm{Tr}(dg g^{-1} A ) - \frac{1}{3} \textrm{Tr} [ (g^{-1} dg)^3]. \cr &&
\end{eqnarray}
The gauge variation of (\ref{CSphi}) is crucial in the anomaly-inflow mechanism, as it can be seen from the descent equation $\delta \mathcal{K}^{i-1}_{2n-i} = d\, \mathcal{K}^{i}_{2n-i-1} $ \cite{Zumino83, Stora83}
\begin{eqnarray} \label{descent}
\delta \int_{M} d\theta \wedge \mathcal{K}^0_{3} &=& \int_{M} d\theta \wedge d \mathcal{K}^1_{2} \cr
&=& - \int_{M} d^2\theta \wedge  \mathcal{K}^1_{2} \cr
&=& - \int_{\Sigma} \mathcal{K}^1_{2}
\end{eqnarray}
where $ \mathcal{K}^0_{3}$ is the Chern-Simons 3-form and connected to $\mathcal{K}^1_{2} $ through the descent equation.
Eqt. (\ref{descent}) relates the four dimensional axial anomaly $d \mathcal{K}^0_{3} $ to the lower-dimensional gauge anomaly $ \mathcal{K}^1_{2} $ on the topological defect $\Sigma$. Note that the phase angle $\theta$ is ambiguous at the origin and hence $d^2\theta$ is singular \cite{Callan-Harvey}
\begin{equation} \label{deltaxy}
d^2\theta = 2 \pi \delta(x) \delta(y) dx \wedge dy
\end{equation}
and we cannot apply $d^2 =0$ in (\ref{descent}).  We will show how this anomaly inflow mechanism (on the string-like topological defects) can be applied to the QCD flux tubes in the next section. Similar applications for the membrane-like topological defects can be found in Refs. \cite{TX1, TX2}, where we study membrane structures in the QCD vacuum.

%%%%%%%%%%%%%%%%%%%%%%%%%%%%%%%%%%%%%%%%%%%%%%
\section{Chromoelectric Flux Tube and Anomaly Inflow}
%%%%%%%%%%%%%%%%%%%%%%%%%%%%%%%%%%%%%%%%%%%%%%

Chromoelectric flux tubes can be generated by quark pair creation, as described in Ref. \cite{Casher}, in the process of hadron production in $e^{+}e^{-}$ annihilation. Given enough energy, a quark pair $q\bar{q}$ is created at some point and then quark and anti-quark move away in opposite direction at a speed $v$ close to the speed of light ($v \approx c$). The chromoelectric flux tube is then formed when the distance between the quark pair becomes $\sim 1 \, \textrm{GeV}^{-1}$ or more. New quark pairs can be produced in a cascade way. The confining constant chromoelectric field inside the flux tube pulls the new quark pairs apart until shorter flux tubes with lower energy form. 

We first consider a simplified case: There already exists a pair of static quark and antiquark $q\bar{q}$ and a chromoelectric flux tube connects the $q\bar{q}$ pair. We also assume that the flux tube is stable again the quantum fluctuations. These are possible scenarios: Firstly the chromoelectric flux tube is treated just like an ordinary electric flux tube, although the color degree of freedom has to be included. This has been studied in details in \cite{Casher} and other references, therefore we focus on another scenario, where the quark condensates are taken into account, and their interactions with quarks are studied as in the usual Nambu-Jona-Lasinio (NJL) effective methods in chiral symmetry breaking and other issues. Just like a magnetic monopole-antimonopole pair creates a string-like region of normal conduction in a superconductor, the quark-antiquark pair creates a region where the chiral quark condensate vanishes, i.e. $\langle \bar{q}q \rangle = 0 $, while outside the chromoelectric flux tube the QCD vacuum has $\langle \bar{q}q \rangle \neq 0 $. This is easy to understand since in sufficiently strong chromoelectric field, the $q\bar{q}$ pairs can be pulled apart and hence the quark condensates disappear. As it will be shown later in this section, it is possible for quark condensates to form a vortex in the chromoelectric flux tubes. This is similar to the phenomenon in condensed matter physics that rotating Bose-Einstein condensates can form  vortices when the angular velocity goes beyond some critical value. 

Using the new variables introduced in Sec. II, the quark part of the QCD action becomes 
\begin{equation}
S_{\textrm{\tiny{quark}}} = \int d^4x [ \bar{\psi} ( i \gamma^{\mu} \tilde{\nabla}_{\mu} - \mathcal{M}_Q ) \psi + g  \bar{\psi} \gamma^{\mu} t_a  \psi B_{\mu}^a ].
\end{equation}
Integrating out $B_{\mu}$ yields the {\it gauged} NJL effective action
\begin{eqnarray}
S_{\textrm{\tiny{gNJL}}} & = & \int d^4x \big( \bar{\psi} ( i \gamma^{\mu} \tilde{\nabla}_{\mu} - \mathcal{M}_Q ) \psi +\int d^4 y ~G(y) \cdot \cr 
&&[ \bar{\psi}(x+y) \Gamma_{A} \psi(x - y)~\bar{\psi}(x-y) \Gamma_{A} \psi(x + y)   ] \big)  \cr
&&
\end{eqnarray}
where $\Gamma_A$ are matrices in Dirac, color and flavor spaces. For simplicity we will only consider one-flavor cases and generalizations to two and more flavors are straightforward. Also the non-local structure of the NJL coupling is not needed in the present study, so we only consider the limiting case where the function $G(x) = G \delta(x)  $ and $G$ is the usual NJL four-fermion coupling constant. The gauged NJL action, which must keep the original symmetry of QCD action, can be parametrized as
\begin{equation} \label{eff}
\mathcal{L}_{\textrm{\tiny{eff}}} = \bar{\psi} \big[ i \gamma^{\mu}(\partial_{\mu}  - i g \tilde{A}_{\mu}) - G ( \sigma +  i \gamma^5 \pi ) + \cdots \big] \psi  
\end{equation}
where the $\sigma$ terms and $\pi$ terms represent the scalar channel and the pseudoscalar channel respectively as in \cite{Hatsuda}.  The ellipsis includes contributions from the vector and pseudovector channels. 
So far this is quite similar to the usual NJL formulation except the {\it gauged} piece, i.e. terms involving the gauge potential $\tilde{A}_{\mu}$. The usual NJL Lagrangian (perhaps even with some extensions) does not lead to a dual superconductor picture, so we have to assume that the $\tilde{A}_{\mu}$-related Abelian dominance lead to quark confinement. This is exactly what we expect from  $\tilde{A}_{\mu}$ -- As mentioned earlier the Wilson loop operator and the Polyakov loop operator only depend on $\tilde{A}_{\mu}$. However we will not study the confinement problem in the present paper. Here we only assume that the chromoelectric field lines are squeezed into tubes and hence the force between a quark-antiquark pair is linear. Let us define a complex field $\Phi$ using the mean fields or condensates $\sigma$  and $\pi$
\begin{equation}
\Phi \equiv \sigma + i \pi = f e^{i \alpha}.
\end{equation}
In order to describe a single straight vortex configuration with winding number $m$, one can take the amplitude and the phase of  $\Phi$ to be, respectively,
\begin{equation} \label{alpha-theta}
f = f(\rho), ~~~~\alpha = m \theta, ~~ m = \pm 1, \pm 2, \cdots 
\end{equation}
For the quark condensates to form a vortex, $\Phi$ should satisfy the boundary condition
\begin{equation}
f(0) = 0, ~ f(\infty) = \textrm{constant}
\end{equation}
and this suggests that we choose
\begin{equation} \label{vortexbc}
|\Phi| (0) = \sqrt{\sigma^2 + \pi^2}\,(0) = 0, ~ |\Phi| (\infty) = \textrm{constant} .
\end{equation}
Can these conditions be satisfied? Following the standard route of computing the effective potential with Schwinger's proper time approach \cite{Schwinger}, one obtains
\begin{eqnarray}
\mathcal{V}_{\textrm{eff}} ( \sigma, \pi)  &=& - \frac{1}{4 G} ( \sigma^2 + \pi^2) - \frac{i }{2} \textrm{Tr} \ln \bigg(  (\partial_{\mu} - i g \tilde{A}_{\mu}) ^2  \cr
&-& \frac{g}{2} \sigma_{\mu\nu} \tilde{F}_{\mu\nu} + \sigma^2 + \pi^2 - i\epsilon \bigg).
\end{eqnarray}
For constant (chromo)electric and/or  (chromo)magnetic background, this effective potential has been calculated before (see \cite{Suganuma} for example). One of the conclusions is that when the chromoelectric field exceeds some critical strength, say $E_{crit} \sim 4 $GeV/fm \cite{Suganuma}, the chiral symmetry is restored inside the chromoelectric flux tube, and we then have a normal phase with chiral symmetry, similar to the normal phase inside the magnetic flux tube of a superconductor. 
Therefore for strong chromoelectric flux tubes, the boundary condition (\ref{vortexbc}) can be satisfied and it is possible for the quark condensates to form a non-trivial topological defect --- a vortex at the same region where the chromoelectric flux tube is located.
Moreover, vortex lattice can be formed in the QCD vacuum just like the Abrikosov lattices in type-II superconductors.
A couple of remarks are in order: First, the boundary condition cannot guarantee the existence of vortex solution, as a trivial configuration with a vanishing winding number is still a possible solution; Second, these vortices emerges at {\it quantum} level, and they are not the classical solutions of the QCD vacuum.

Now let us study the effective Lagrangian (\ref{eff}) in the background of (\ref{vortexbc}). We focus on the vortices with winding number $m= \pm 1$ in the quark condensates. 
From Sec. II  the gauge potential $\tilde{A}_{\mu}$ can be rewritten as 
\begin{equation}
\tilde{A}_{\mu} = \hat{\Omega}^{\textrm{\tiny{vac}}}_{\mu} +  Z_{\mu} {\bf n}
\end{equation}
where $Z_{\mu}  \equiv A_{\mu} \cdot {\bf n} + H_{\mu} $ is the $U(1)$ potential introduced in eqt. (\ref{U1}) 
and the corresponding field strength
\begin{equation}
\tilde{F}_{\mu\nu} = Z_{\mu\nu} {\bf n}
\end{equation}
\begin{equation}
Z_{\mu\nu} = \partial_{\mu} Z_{\nu} - \partial_{\nu} Z_{\mu} 
\end{equation}
which shows the {\it abelian} character of $Z_{\mu}$.  Here we have dropped the last term in eqt. (\ref{Hmunu}) since in our case ${\bf n}$ can be well-defined in the flux tube with a pair of color sources at its ends, not like the non-physical Dirac string. In the flux tube one can simply impose a gauge on ${\bf n}$, say ${\bf n} = (0,0,1)$ in the color space as in \cite{Cho, Kondo, Shabanov}.  In fact as it is pointed out in \cite{Shabanov, Cho}, if one sets ${\bf n} = (0,0,1)$ in the entire space,  eqt. (\ref{reduction}) will turns into the maximal Abelian gauge condition and the gauge transformation needed to reach this gauge will contain singularities or defects with the quantum numbers of magnetic monopoles. More complicated gauge conditions on $ {\bf n}$ can be associated with the knot topology of the QCD vacuum \cite{Cho2, Langmann:1999nn}. 

Now we choose a longitudinal variation of gauge potential with respect to $ \hat{\Omega}^{\textrm{\tiny{vac}}}_{\mu} $
\begin{equation} \label{Az}
Z_{\mu} = ( Z_{0}, Z_{1}, 0, 0),
\end{equation}
for example, $Z_{\mu} = ( 0, - E t, 0, 0)$ corresponds to a constant electrical field $\vec{E}$ background along the $x_1$-axis (or the $z$-axis) . We will consider more general configurations in the next section. Writing the quark field variation $\psi$ in terms of $\psi_{R,L} = 1/2 \,(1 \pm r^5) \,\psi$, the Dirac equation splits into two equations, similar to Ref.\cite{Callan-Harvey}
\begin{eqnarray}  \label{Dirac4D}
i \gamma^i (\partial_i - ig Z_i) \psi_L &+& i( \gamma^2 \cos \theta  + \gamma^3 \sin \theta ) \partial_{\rho} \psi_L \cr &+& f(\rho) e^{- i \theta} \psi_R =0, \cr
i \gamma^i (\partial_i - ig Z_i) \psi_R &+& i( \gamma^2 \cos \theta  + \gamma^3 \sin \theta ) \partial_{\rho} \psi_R \cr &+& f(\rho) e^{+ i \theta} \psi_L =0, 
\end{eqnarray}
where $i = 0, 1$. From these two equations one can derive the equation of motion for a two-dimensional spinor $\chi_{L} (x_0, x_1)$
\begin{equation} \label{Dirac2D}
i \gamma^i (\partial_i - ig Z_i) \chi_{L} = 0
\end{equation}
and $\chi_{L}$ satisfies $ (-r^0 r^1) \chi_{L} = - \chi_{L} $. Then solution to (\ref{Dirac4D}) obtains the same exponential profile as in Ref.\cite{Callan-Harvey}
\begin{equation} \label{4Dpsi}
\psi_L = \chi_{L} \, \exp \big[- \int_0^{\rho} f(\rho') d\rho' \big]
\end{equation}
and $\psi_R = i \gamma^2 \psi_L $. Note that the difference between equations (\ref{Dirac4D}), (\ref{Dirac2D}) and those in Ref.\cite{Callan-Harvey} is that we have included the gauge potential $Z_{\mu}$ explicitly. The exponential profile in (\ref{4Dpsi}) shows that chiral zero modes are localized on the flux tube. However, let us keep in mind that this is based on the assumption (\ref{Az}). Other gauge configurations could modify the exponential profile significantly, indicating the transverse motion of quark zero modes. Note that there is a two-dimensional chirality defined by the $-\gamma^0 \gamma^1$ which is the two-dimensional analogue of $\gamma^5$ in four dimensions. From the eqt. (\ref{Dirac2D}) we see that chiral zero modes are coupled to a gauge field. It is well known that a gauge anomaly appears in such a situation \cite{Callan-Harvey, Naculich}
\begin{equation}
\mathcal{D}^k J_k = \frac{1}{2 \pi} \epsilon^{ij} \partial_i Z_j, ~~~~ i,j,k = 0,1.
\end{equation}
This seems to be a problem since the original theory QCD should not have any gauge anomaly in the four dimensional spacetime. This paradox, similar to what we have encountered in the axion string model in Sec. III, can be solved if we apply the anomaly inflow mechanism \cite{Callan-Harvey} to the case of QCD flux tubes. The key is to realize that massive modes of the Dirac equation which live off the vortex mediate an effective interaction between the quark condensates and the chromoelectric field in the flux tube. The induced vacuum current can be calculated as in \cite{Callan-Harvey, Naculich}
\begin{equation} \label{induc}
J^{\textrm{\tiny{ind}}}_{\mu} =  \frac{1}{8 \pi^2} \epsilon_{\mu\nu\rho\tau} Z^{\rho\tau} \partial^{\nu} \theta
\end{equation}
which can be converted to an effective action
\begin{equation} \label{csEff}
\mathcal{L}_{\textrm{\tiny{c-s}}} = -\frac{1}{8 \pi^2}\int d^4 x \, \partial_{\mu} \theta \, K^{\mu}
\end{equation}
where the Chern-Simons current is 
\begin{equation}
K_{\mu} \equiv  \epsilon_{\mu\nu\rho\tau} Z^{\nu} Z^{\rho\tau} 
\end{equation}
It is easy to see that the gauge variation of $\mathcal{L}_{\textrm{\tiny{c-s}}}$ cancels precisely the gauge anomaly on the vortex configuration.  The consequence is that a topological effective action (\ref{csEff}) emerges in the bulk and should be included in the phenomenological applications of the flux-tube models. An intriguing fact is that although the Chern-Simons current coupling in (\ref{csEff}) yields the topological charge term (\ref{TC}) after integration by parts, it does not mean that they are equivalent since they have different gauge transformation laws.  They could be related if one includes the boundary terms and treats them properly, or promotes the usual $\theta$-parameter to a local variable which transforms under the gauge transformation. In Ref. \cite{TX1} the $\theta$-variable is interpreted as a compactified Ramond-Ramond potential from the holographic QCD point of view.  In Ref. \cite{TX2} it indicates codimension-one membrane-like topological charge structures in the QCD vacuum, without reference to the string theory framework. In the present arena it might be possible to reconcile two expressions of topological terms by gradually increasing the radius of the flux tube, from infinitesimal to some finite value, however we defer the question about their connections to future work.

We have shown that a strong chromoelectric flux tube may induce a vortex configuration in the quark condensates. The example we studied above has winding number $m=\pm 1$. Given more energy, which means stronger color electric field strength, vortex with larger winding number or more vortices can be created as well.
We have also shown that quark zero modes located at the vortex become chiral, which leads to gauge anomaly in the two-dimensional defect. This can only be cancelled by a topological term containing Chern-Simons current in the bulk of the four-dimensional spacetime. Therefore it relates the gauge anomaly in the vortex to the chiral anomaly in the bulk. 
From the current inflow point of view one may consider the flux tube as an {\it ``anomaly battery"}. It is ``charged" by the current  (\ref{induc}) whose only non-vanishing component is $J_{\rho} = g E / (4 \pi \rho)$ (see Fig 1). It is interesting that this current flows in the direction perpendicular to the chromoelectric field $E$ like that of the Hall effect, as first noticed by Callan and Harvey \cite{Callan-Harvey}.    

\begin{figure}[!h]
\begin{center}
    \includegraphics[height=45mm, width=60mm]{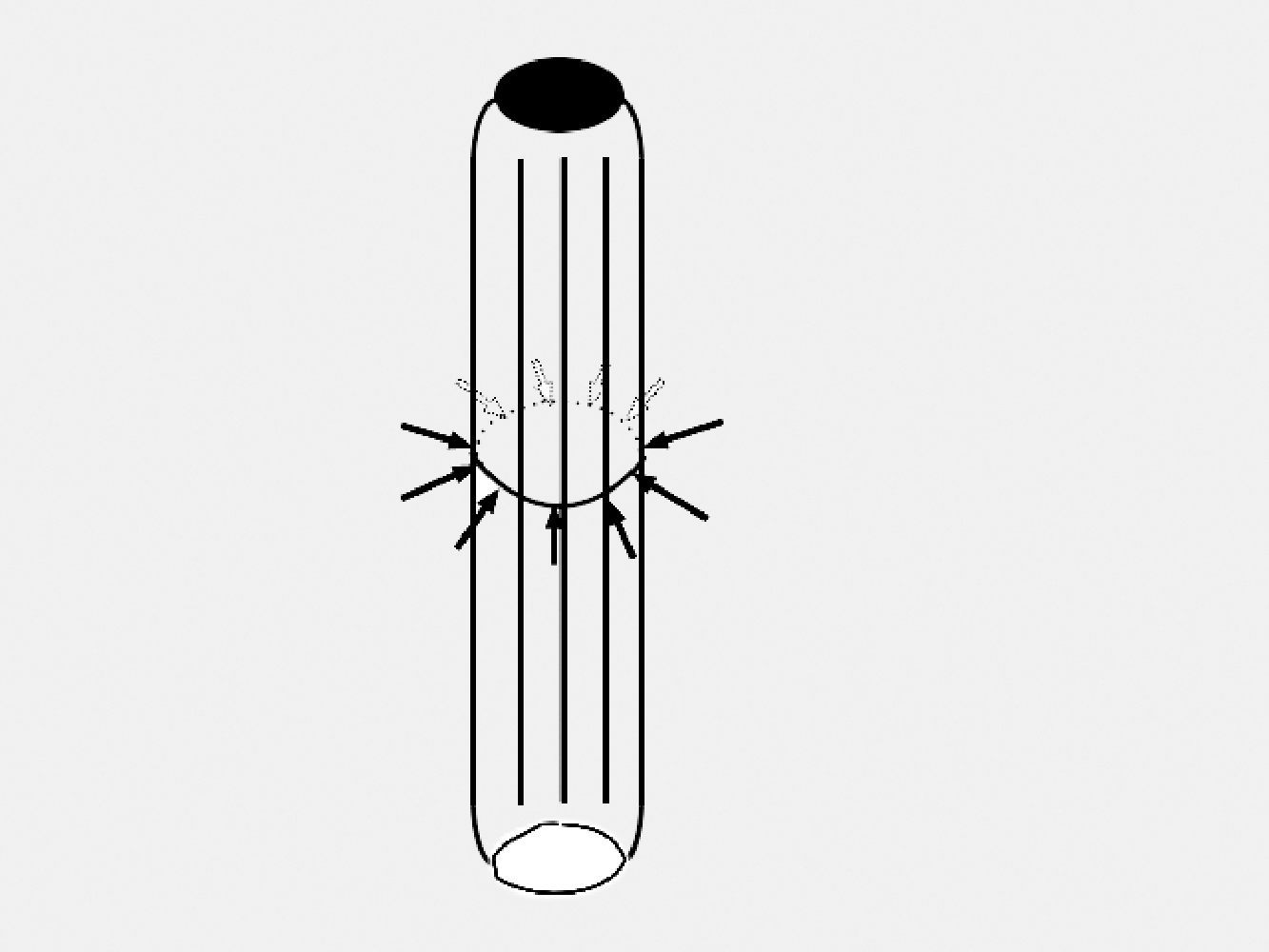}
    \caption{The anomaly-inflow picture of a chromoelectric flux tube in the QCD vacuum. The black circle and the white circle represent color sources (e.g. a quark-antiquark pair) and the lines between them are chromoelectric field lines squeezed into a thin tube in the vacuum (shaded area). The arrows on the circle indicate the directions of the induced vacuum current on a particular cross-section of the flux tube, similar to the configuration of the Hall effect. }
\end{center}
\label{fluxtubeInflow}
\end{figure}

\section{Flux Tube with Magnetic Fields and Chiral Magnetic Effect}

In this section we extend our formulation to the flux tubes with extra magnetic fields (chromomagnetic and electromagnetic). Such field configurations can be applied to some interesting phenomena, like the glasma state \cite{Lappi06} and the chiral magnetic effect \cite{Fukushima08, Kharzeev07}. 
Also the quark creation in such gauge configuration is different from those chirality-conserving processes which are the usual pair creation in strong (chromo)electric fields \cite{Casher} with the Schwinger pair-creation mechanism \cite{Schwinger}, where anomaly does not seem to play a role. Let us take quark-gluon-plasma (QGP) into account. The formation of QGP also involves color flux tubes. During the ultra-relativistic collision of heavy nuclei, net color charges develop and consequently color flux tubes connect these color charges. Quark-antiquark and gluon pairs are produced from the color fields via the Schwinger mechanism as mentioned before. Nevertheless pair creation can be enhanced when the chromomagnetic field presents and has a non-vanishing component in the direction of the chromoelectric field, more precisely, $\vec{E} \cdot \vec{B} \neq 0$. For example, in \cite{Lappi06} it has been proposed that longitudinal color electric and magnetic fields are produced immediately after the high energy hadronic collisions.
Such gauge configuration has non-vanishing winding number and can separate charge in a background (electromagnetic) magnetic field. Consequently an electromagnetic current is generated along this magnetic field. 
This is dubbed as the chiral magnetic effect \cite{Fukushima08, Kharzeev07} where chiral anomaly plays an important role.

There are different approaches in understanding chiral anomalies, e.g. perturbation theory,  path integral and index theorem. The ``Dirac sea" explanation in \cite{Nielsen83} is appealing and particularly suitable for our case. 
From eqt. (\ref{Dirac2D})
\begin{equation} \label{Dirac2DL}
i \gamma^i (\partial_i - ig Z_i) \chi_{L} = 0,
\end{equation}
with $Z_{\mu} = (0, -E(t), 0, 0) $ it becomes
\begin{equation}
\frac{\partial \chi_L}{\partial t} = - \frac{\partial \chi_L}{\partial z} - i g E(t) \chi_L 
\end{equation}
For a uniform chromoelectric field $E(t) = E \,t$, this reduces to the case studied in \cite{Nielsen83}, where
a very simple argument leads to the axial anomaly: the creation/annihilation rate of right/left-handed particles per unit time and unit length depends on the change of the fermi surface. According to \cite{Nielsen83} the annihilation rate of the left-handed particles is
\begin{equation} \label{ABJ}
\frac{\partial n_L}{\partial t} \equiv \dot{n}_L =  - \frac{g}{2 \pi} E
\end{equation} 
based on the dispersion relation from (\ref{Dirac2DL}) $\omega(p) = - p$ and the classical equation of a charged particle in a uniform electric field $\partial p /\partial t = g E $. Therefore eqt. (\ref{ABJ}) gives the axial anomaly for Dirac particles in two-dimensional spacetime 
\begin{equation}
\dot{n}_R - \dot{n}_L =  \frac{g}{\pi} E
\end{equation}
which is consistent with the relation usually obtained from the index theorem
\begin{equation}
n_R - n_L = -\frac{g}{2 \pi} \int d^2 x \, \epsilon_{ij} Z^{ij}
\end{equation}
This can be generalized to the four-dimensional spacetime where we add a constant chromomagnetic field in the $z$-direction. In the symmetric gauge $Z_{\mu} = (0, -E \, t, -\frac{1}{2}B y, \frac{1}{2}B x, 0) $, the Dirac equation in the collinear color electric and magnetic fields becomes
\begin{eqnarray}  \label{EB_Dirac4D}
i \gamma^i (\partial_i - ig Z_i) \psi_L &+& i( \gamma^2 \cos \theta  + \gamma^3 \sin \theta ) \partial_{\rho} \psi_L \cr &+& \frac{1}{2}g B \rho ( \gamma^3 \cos \theta - \gamma^2 \sin \theta) \psi_L \cr
&+& f(\rho) e^{- i \theta} \psi_R =0, \cr
i \gamma^i (\partial_i - ig Z_i) \psi_R &+& i( \gamma^2 \cos \theta  + \gamma^3 \sin \theta ) \partial_{\rho} \psi_R \cr &+&\frac{1}{2}g B \rho ( \gamma^3 \cos \theta - \gamma^2 \sin \theta) \psi_R \cr
&+& f(\rho) e^{+ i \theta} \psi_L =0, 
\end{eqnarray}

The profile of the quark zero modes now has an extra $B$-dependent factor
\begin{equation} \label{4DpsiB}
\exp \big[- \int_0^{\rho} f(\rho') d\rho'   - \frac{1}{4} g B \rho^2 \big]
\end{equation}
(More general cases like time-dependent $E(t), B(t)$ fields can be studied in a similar way) and the direction of color magnetic field in the $z$-axis also matters ($\vec{B}$ parallel or anti-parallel to $\vec{E}$). If the $f(\rho)$ factor is set to zero, the equations in (\ref{EB_Dirac4D}) will reduce to the Dirac equation in constant electric field and magnetic field background, which has been studied before (see e.g. \cite{Nikishov}). 
In general we cannot treat the configuration as a string and must solve the equation in the four-dimensional spacetime.  However, it is easy to see that if color electric and magnetic fields are (anti)parallel, the topological charge density 
\begin{equation}
\frac{1}{4} Z_{\mu\nu} \tilde{Z}^{\mu\nu} = \vec{B} \cdot \vec{E} \neq 0
\end{equation}
and the winding number of a gauge configuration is
\begin{equation} \label{EBQ}
Q =  \frac{g^2 N_f}{32 \pi^2} \int d^4 x  Z_{\mu\nu} \tilde{Z}^{\mu\nu} \neq 0,
\end{equation}
hence the chirality is not conserved
\begin{equation} \label{RLQ}
{N}_R - {N}_L = -2 N_f Q \neq 0.
\end{equation}
Now we include the electromagnetic fields. Inside the flux tube, it is a deconfinement phase with restored chiral symmetry; the localized quarks are massless and these are the same conditions as those considered in the chiral magnetic effect \cite{Fukushima08, Kharzeev07}. 
For the additional (electro)magnetic background we need add an electromagnetic potential  $A_{\mu}^{\textrm{\tiny{M}}}$ to the equation (\ref{EB_Dirac4D}). Instead of solving the equation, we study the chiral magnetic effect through the anomaly argument as follows: 
From the equations (\ref{EBQ}, \ref{RLQ}) one notices that there should be net chiral or anti-chiral quark zero modes in the flux tube with both chromoelectric and chromomagnetic fields; Since these (anti)chiral zero modes are coupled to the electromagnetic potential $A_{\mu}^{\textrm{\tiny{M}}}$, they lead to the gauge anomaly with respect to the electromagnetic potential and the anomaly-inflow mechanism should apply in the same way. Following the procedure in the previous section, we obtain an extra effective coupling in the bulk
\begin{equation} \label{mcs}
\Delta\mathcal{L}_{\textrm{\tiny{MCS}}} \sim -\frac{e^2}{8 \pi^2} \, \partial_{\mu} \alpha \, K^{\mu}_{\textrm{\tiny{MCS}}}
\end{equation}
where $ K^{\mu}_{\textrm{\tiny{MCS}}} =  \epsilon^{\mu\nu\rho\tau} A_{\nu} F_{\rho\tau} $ is the Maxwell-Chern-Simons current, and we have also replaced the $\theta$-variable by a more general phase $\alpha$ (see eqt. (\ref{alpha-theta})). This is because the flux tubes that we have studied so far are stationary, whereas the real ones should be dynamical in the sense that they may rotate, vibrate, or even decay into particles  \cite{Kharzeev:2001vs, Shuryak:2002qz, Lappi06}. By the spacetime dependence and the Lorentz symmetry, the $\theta$-variable in the stationary case should be promoted to a spacetime-dependent phase $\alpha(\vec{x},t)$. Eqts (\ref{EBQ}) and (\ref{mcs}) suggest that there is an interplay between the Abelian anomaly and the non-Abelian anomaly. Consequently the Maxwell equation becomes
\begin{equation}
\partial_{\mu} F^{\mu\nu} = J^{\nu} - \frac{e^2}{2 \pi^2} \, \partial_{\mu} \alpha \, \tilde{F}^{\mu\nu}
\end{equation}
which yields the component equation
\begin{equation}
\nabla \times \vec{B}^{\textrm{\tiny{M}}} - \frac{\partial \vec{E}^{\textrm{\tiny{M}}}}{\partial t} = \vec{J} + \frac{e^2}{2 \pi^2} \,( \dot{\alpha} \vec{B}^{\textrm{\tiny{M}}} - \nabla \alpha \times \vec{E}^{\textrm{\tiny{M}}}).
 \end{equation}
Taking the limit $\vec{E}^{\textrm{\tiny{M}}} \rightarrow 0$ as in \cite{Fukushima08, Kharzeev07}, and noticing that $\nabla \times \vec{B}^{\textrm{\tiny{M}}}=0$, one
obtains
\begin{equation} \label{JB}
\vec{J} \sim -\dot{\alpha} \vec{B}^{\textrm{\tiny{M}}}
\end{equation}
which is the induced current due to the chiral magnetic effect. The coefficient $\dot{\alpha}$ can be identified as the chiral chemical potential \cite{Fukushima08, Kharzeev07} and $\nabla \alpha$ is a superflow term suppressed by the vanishing electric field $\vec{E}^{\textrm{\tiny{M}}}$.
This simple derivation is based on our particular flux-tube model and the anomaly-inflow mechanism. More general and rigorous derivations can be found in \cite{Fukushima08, Kharzeev07}.

%\newpage

\section{Discussions and Conclusions}

The effective Chern-Simons coupling, $\partial_{\mu}\theta \, K^{\mu}_{\textrm{\tiny{c-s}}}$ obtained from the QCD flux tube model and the anomaly inflow mechanism, can be put in the familiar form $\theta F \tilde{F}$ after integration by parts. This suggests the emergence of the chiral charge in the bulk.  Our formulation explains why such topological charge is needed, but does not explain how it is distributed in the QCD vacuum. Instantons, center vortices and domain walls are possible gauge configurations which can ``recharge" the ``anomaly battery" --- the flux tube. Also the appearance of this topological term is associated with the quark-creation, and it could be related to the instanton or sphaleron cases in \cite{Kharzeev:2001vs, Shuryak:2002qz, Lappi06} where the production of particles is due to the decay of Chern-Simons charge. 

We close by pointing out that applying the anomaly-inflow mechanism to the QCD flux tubes should not depend on particular formulations. Even for the Cho-Faddeev-Niemi decomposition that we used in this paper, there is a dual formulation \cite{Faddeev, Langmann:1999nn}
\begin{equation}
\tilde{A}_{\mu} = - C_{\mu} {\bf n} - g^{-1} {\bf n} \times \partial_{\mu} {\bf n}  + \phi_1 \, \partial_{\mu} {\bf n} - \phi_2 \, {\bf n} \times \partial_{\mu} {\bf n} 
\end{equation}
where $\phi_{1,2}$ are two real scalar fields which are dual to the color vector ${\bf n}$. Given proper vacuum expectation values for these fields, a (dual) abelian-Higgs model can be constructed \cite{Langmann:1999nn}. It would be interesting to see how to use these variables to formulate the color flux tubes. Note that these variables appear at the classical level while some dynamical degrees of freedom in our model emerge at the quantum level. 

To summarize, we used a set of new QCD variables from the Cho-Faddeev-Niemi decomposition to formulate QCD flux-tube models and applied the anomaly-inflow mechanism to the studies of chromoelectric flux tubes, quark (pair)-creation and chiral magnetic effect. At quantum level, quark condensates in the QCD vacuum can induce vortex-like structures in a chromoelectric flux tube. Quark zero modes trapped in the vortex are chiral and generate two-dimensional gauge anomaly. To cancel this anomaly a Chern-Simons type effective action is introduced, hence a topological charge density term naturally appears in the bulk spacetime. This scenario can also be used to study anomaly enhanced quark-pair creation, chiral magnetic effect and etc. In this paper we only considered straight string-like color flux tubes. It is amusing to study more complicated situations, like the $\Delta$-shape and Y-shape flux-tube configurations for baryons.

\acknowledgments

The author thanks Hank Thacker, Peter Arnold and Kerson Huang for discussions in QCD and quark-gluon plasma. He also thanks Peter Minkowski for a careful reading of the manuscript and for his comments and suggestions, as well as the anonymous referee for constructive comments. This work is supported by the research funds from the Institute of Advanced Studies, Nanyang Technological University, Singapore.

\end{document}